\documentclass[12pt]{article}
\usepackage{graphicx}
\usepackage{amssymb}
\usepackage{amsmath}
\usepackage{subfigure}

\def\BsKK{B_s\to K^+ K^-}
\def\Amix{\mathcal{A}^{\rm mix}_{\rm CP}}

\begin{document}

\thispagestyle{empty}

\begin{flushright}
Nikhef-2010-043\\
\end{flushright}

\vspace{1.0truecm}
\begin{center}
\boldmath
\large\bf A Fresh Look at $B_{s,d}\to \pi\pi,\pi K, KK$ Decays
\unboldmath
\end{center}

\vspace{0.3truecm}
\begin{center}
Robert Fleischer  \,and\, Robert Knegjens\\[0.1cm]
{\sl Nikhef, Science Park 105, 
NL-1098 XG Amsterdam, The Netherlands}
\end{center}

\vspace{0.3truecm}

\begin{center}
{\bf Abstract}
\end{center}

{\small
\vspace{0.2cm}\noindent
Using updated measurements and $SU(3)$-breaking form factors, we have a 
detailed look at the $B_d\to\pi^+\pi^-$, $B_s\to K^+K^-$ and $B_d\to\pi^\mp K^\pm$, 
$B_s\to \pi^\pm K^\mp$ systems. The corresponding decays are related to each other
by the $U$-spin symmetry of strong interactions and offer determinations of the angle
$\gamma$ of the unitarity triangle. In the former case, we obtain 
$\gamma=\left(68.3^{+4.8}_{-5.7}\mbox{}^{+5.0}_{-3.7}\right)^\circ$, which is in excellent 
agreement with the Standard-Model fits of the unitarity triangle. The first errors 
correspond to experimental input uncertainties, while the latter are an estimate
of $U$-spin-breaking effects. In view of this result, large CP-violating new-physics 
effects at the amplitude level are excluded. However, the effective $B_s\to K^+K^-$
lifetime and the mixing-induced CP violation in this channel offer interesting probes 
for New Physics in $B^0_s$--$\bar B^0_s$ mixing. In the case of the $B_d\to\pi^\mp K^\pm$, 
$B_s\to \pi^\pm K^\mp$ system, using additional information from $B^\pm\to \pi^\pm K$, 
we obtain a bound of  $\gamma\leq \left(71.8^{+5.4}_{-4.3}\right)^\circ$, and the range
$24^\circ \leq \gamma \leq 71^\circ$. We perform also tests of the $U$-spin symmetry 
and do not find any indication for large non-factorizable corrections.}

\vspace{1.2truecm}

\begin{center}
{\sl Talk at the 6th International Workshop\\ on the CKM Unitarity Triangle
(CKM2010)\\
Warwick, United Kingdom, 6--10 September 2010\\
To appear in the Proceedings}
\end{center}

\vfill
\noindent
December 2010

\newpage
\thispagestyle{empty}
\vbox{}
\newpage
 
\setcounter{page}{1}

\section{Introduction}
The key problem in the analysis of non-leptonic $B$ decays is related to the hadronic
matrix elements of local four-quark operators. A powerful method to deal with this challenge
is offered by the flavour symmetry of strong interactions, which allows us to get the 
relevant hadronic parameters from experimental data. In this respect, the $U$-spin-related decays 
$B_d\to\pi^+\pi^-$, $B_s\to K^+K^-$ are particularly interesting, allowing a determination 
of the angle $\gamma$ of the unitarity triangle (UT) \cite{RF-BsKK}. The advantage with respect
to conventional $SU(3)$ strategies is twofold: no additional dynamical assumptions, which 
could be spoiled by large rescattering effects, have to be made, and electroweak penguins
are automatically included. The relevant observables are the CP-averaged branching ratios,
and the direct and mixing-induced CP asymmetries:
\begin{equation}
\frac{\Gamma(B^0_q(t)\to f)-
\Gamma(\bar B^0_q(t)\to f)}{\Gamma(B^0_q(t)\to f)+
\Gamma(\bar B^0_q(t)\to f)}\propto {\cal A}_{\rm CP}^{\rm dir}\cos(\Delta M_q t)+
{\cal A}_{\rm CP}^{\rm mix}\sin(\Delta M_q t).
\end{equation}
There is another interesting $U$-spin-related pair, which is given by 
$B_d\to\pi^\mp K^\pm$, $B_s\to \pi^\pm K^\mp$ \cite{GR}. By using further input
from $B^\pm\to \pi^\pm K$, $\gamma$ can be extracted. 

A comprehensive analysis of these decays was performed in 2007 \cite{RF-Bhh}. 
Here we will have a fresh look at this topic, using updated measurements from CDF 
\cite{CDF-BsKK} and data taken by Belle at $\Upsilon(5S)$ \cite{Belle}, 
as well as updated information on $SU(3)$-breaking form factors \cite{DuMe}. 
For a detailed discussion, including derivations of the relevant formulae, further 
numerical results and references, the reader is referred to Ref.~\cite{FK}.

\section{\boldmath The $B_d\to \pi^+\pi^-$, $B_s\to K^+K^-$ System\unboldmath}
In the Standard Model (SM), the $B_s^0\to K^+K^-$ and $B^0_d\to\pi^+\pi^-$ decay 
amplitudes can be written as follows \cite{RF-BsKK}:
\begin{equation}\label{ampl}
A(B_d^0\to\pi^+\pi^-)\propto {\cal C}
\left[e^{i\gamma}-d\,e^{i\theta}\right], \quad
A(B_s^0\to K^+K^-)\propto {\cal C}'\left[e^{i\gamma}+\frac{1}{\epsilon}d'e^{i\theta'}\right], 
\end{equation}
where $\lambda\equiv|V_{us}|$, 
$\epsilon\equiv\lambda^2/(1-\lambda^2)$ are CKM factors, while ${\cal C}^{(\prime)}$ and 
$d^{(\prime)} e^{i\theta^{(\prime)}}$  are CP-conserving hadronic parameters that describe,
loosely speaking, tree contibutions and the ratio of penguin to tree amplitudes, respectively. 
The $U$-spin symmetry implies 
\begin{equation}\label{rel-1}
d'=d, \quad \theta'=\theta. 
\end{equation}
Since $de^{i\theta}$ and $d'e^{i\theta'}$ are actually ratios of 
hadronic amplitudes, $U$-spin-breaking form factors and decay constants cancel. 
On the other hand, decay constants and form factors do not cancel in $|{\cal C}'/{\cal C}|$.
Consequently, they enter also the observable
\begin{equation}\label{K-def}
K\propto \frac{1}{\epsilon}\,\left|\frac{{\cal C}}{{\cal C}'}\right|^2
\left[\frac{\mbox{BR}(B_s\to K^+K^-)}{\mbox{BR}(B_d\to\pi^+\pi^-)}\right]=
\frac{1}{\epsilon^2}\left[\frac{\epsilon^2+2\epsilon d'\cos\theta'\cos\gamma+
d'^2}{1-2d\cos\theta\cos\gamma+d^2}\right]
\stackrel{\rm exp}{=}51.8^{+12.7}_{-15.0},
\end{equation}
where the numerical value uses the updated QCD sum-rule calculation of Ref.~\cite{DuMe}.

With the amplitude parameterizations in (\ref{ampl}),  
we can express the CP asymmetries of these decays as functions of $d^{(\prime)}$, 
$\theta^{(\prime)}$ and $\gamma$. In the case of the mixing-induced CP asymmetries,
also the $B^0_q$--$\bar B^0_q$ mixing phases $\phi_q$ ($q\in\{d,s\}$) enter;  
$\phi_d = (42.4^{+3.4}_{-1.7})^\circ$ is already known \cite{FFJM}, and $\phi_s$ can be
determined through $B^0_s\to J/\psi \phi$. The measured values of the CP asymmetries
can then be converted into theoretically clean contours in the $\gamma$--$d^{(\prime)}$
planes. Using the $U$-spin relation $d=d'$, $\gamma$ as well as $\theta$ and $\theta'$
can be determined, where the strong phases offer a test of the $U$-spin
symmetry \cite{RF-BsKK}. 

\begin{figure}[t]
\centering
\subfigure[]{
	\includegraphics[width=7truecm]{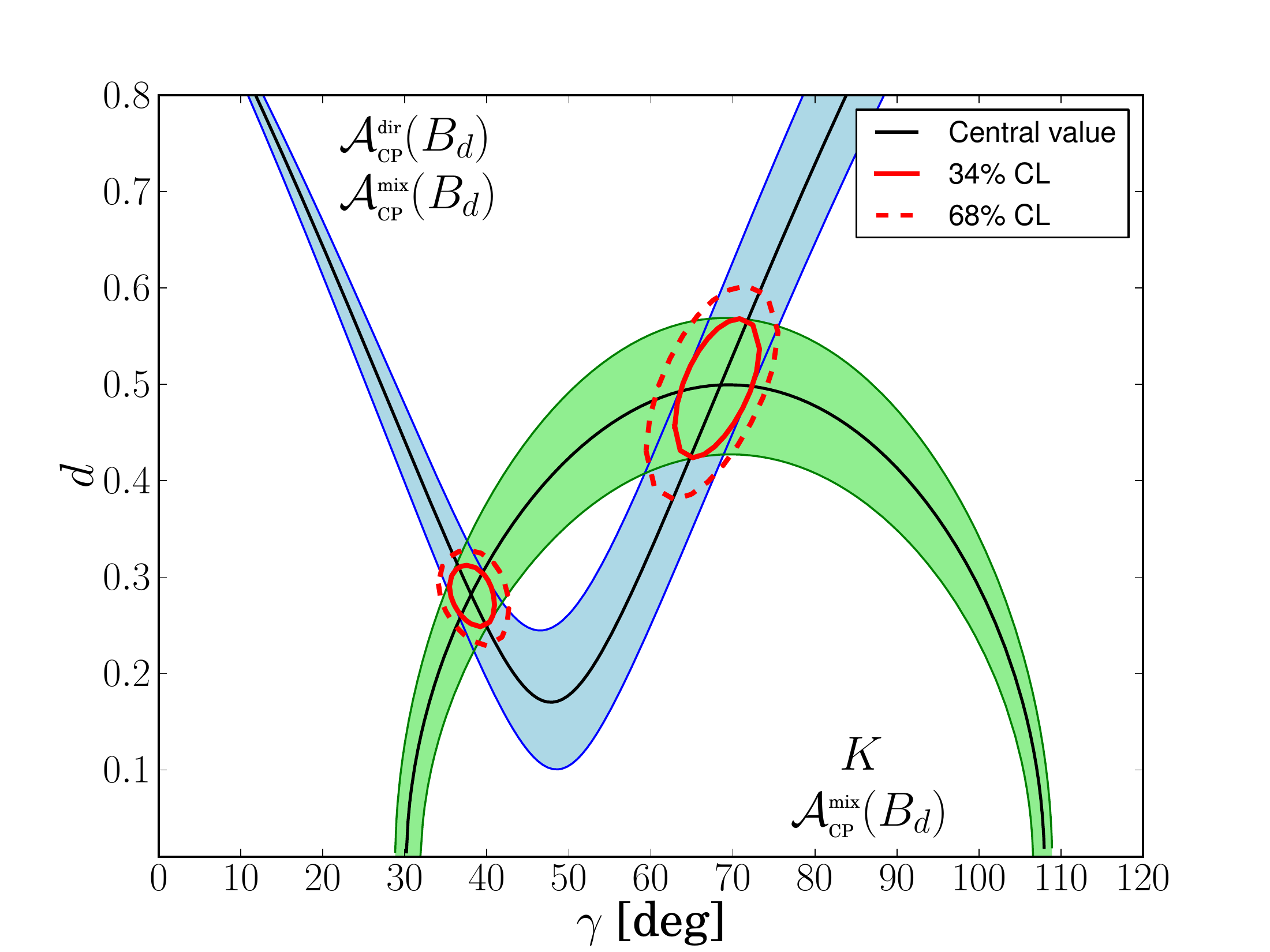}\label{fig:current}
}
\subfigure[]{
	\includegraphics[width=7truecm]{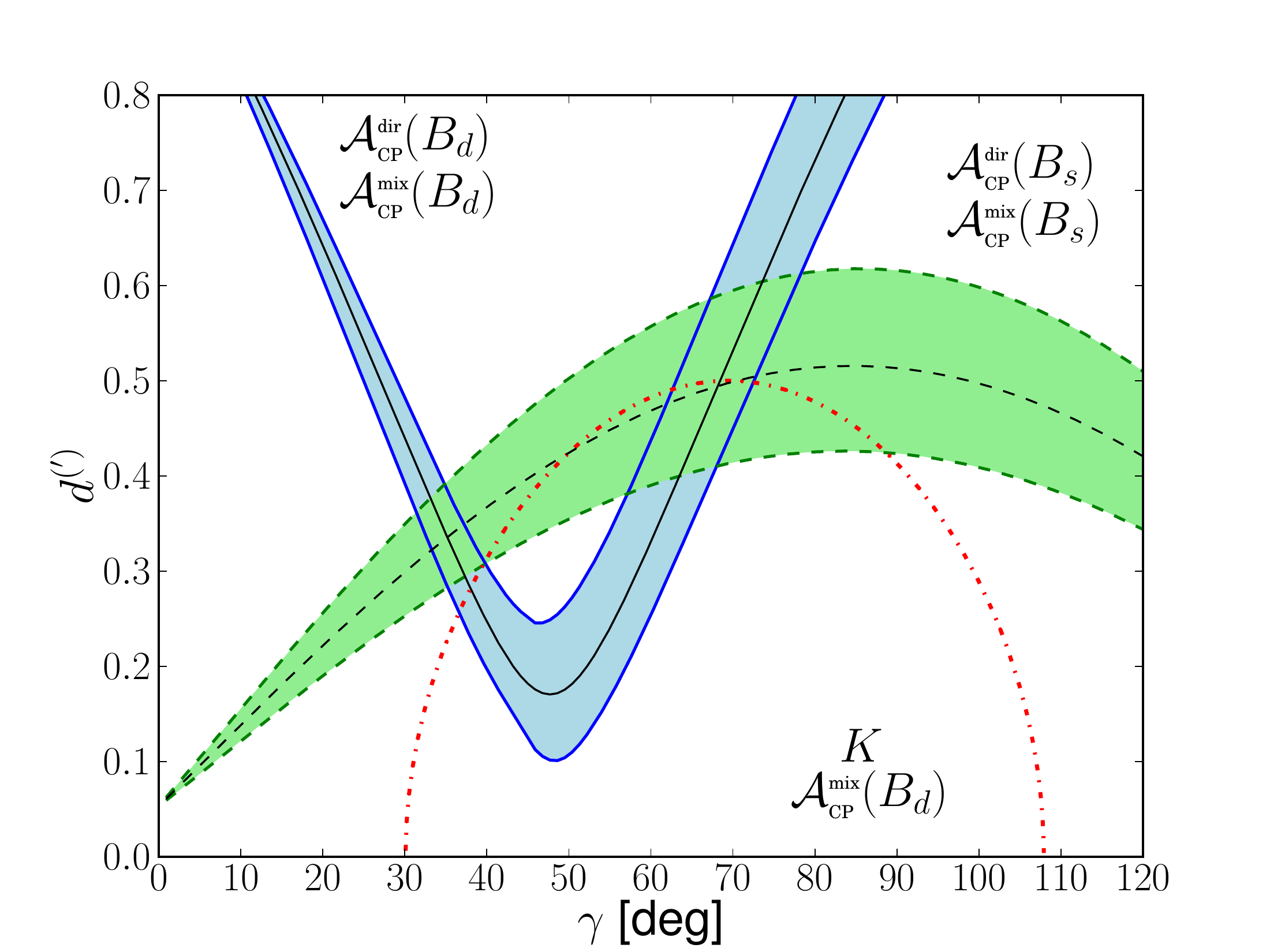}\label{fig:optimal}
}
   \caption{Contours for the determination of $\gamma$ from the 
   $B_d\to\pi^+\pi^-$, $B_s\to K^+K^-$ system: {(a)} $1\,\sigma$ error 
   bands and 68\% C.L. regions for current data; {(b)} 
    illustration of the optimal determination of $\gamma$ 
    using also CP violation in $B_s\to K^+K^-$ (SM case).}
\end{figure}

So far, only $\mbox{BR}(B_s\to K^+K^-)$ has been measured by CDF \cite{CDF-BsKK}
and Belle \cite{Belle}. Unfortunately, data on the CP-violating effects in this channel 
are not yet available. In the case of $B_d\to\pi^+\pi^-$, there is good agreement between 
the BaBar and Belle results for the mixing-induced CP asymmetry; the Heavy Flavour
Averaging Group (HFAG) gives the average 
${\cal A}_{\rm CP}^{\rm mix}(B_d\to\pi^+\pi^-)=0.65\pm0.07$. On the other
hand, we are unfortunately still facing a discrepancy between the BaBar and Belle measurements for the direct CP asymmetry, which, in our opinion, makes it problematic to combine them 
in an average. This feature is reflected by the averages  in the literature: HFAG gives 
${\cal A}_{\rm CP}^{\rm dir}(B_d\to\pi^+\pi^-)=-0.38 \pm 0.06$, whereas the 
Particle Data Group quotes the same central value with a larger error of 0.17. 
In view of this unsatisfactory situation, we would like to avoid these averages, utilizing instead 
the direct CP-asymmetry measurement of the $SU(3)$-related $B_d\to \pi^\mp K^\pm$ channel,
which is well settled. This gives ${\cal A}_{\rm CP}^{\rm dir}(B_d\to\pi^+\pi^-)= -0.26\pm0.10$, 
where we have generously inflated the error to allow for possible $SU(3)$-breaking corrections.

\begin{figure}[t]
   \centering
   \subfigure[]{
	   \includegraphics[width=7truecm]{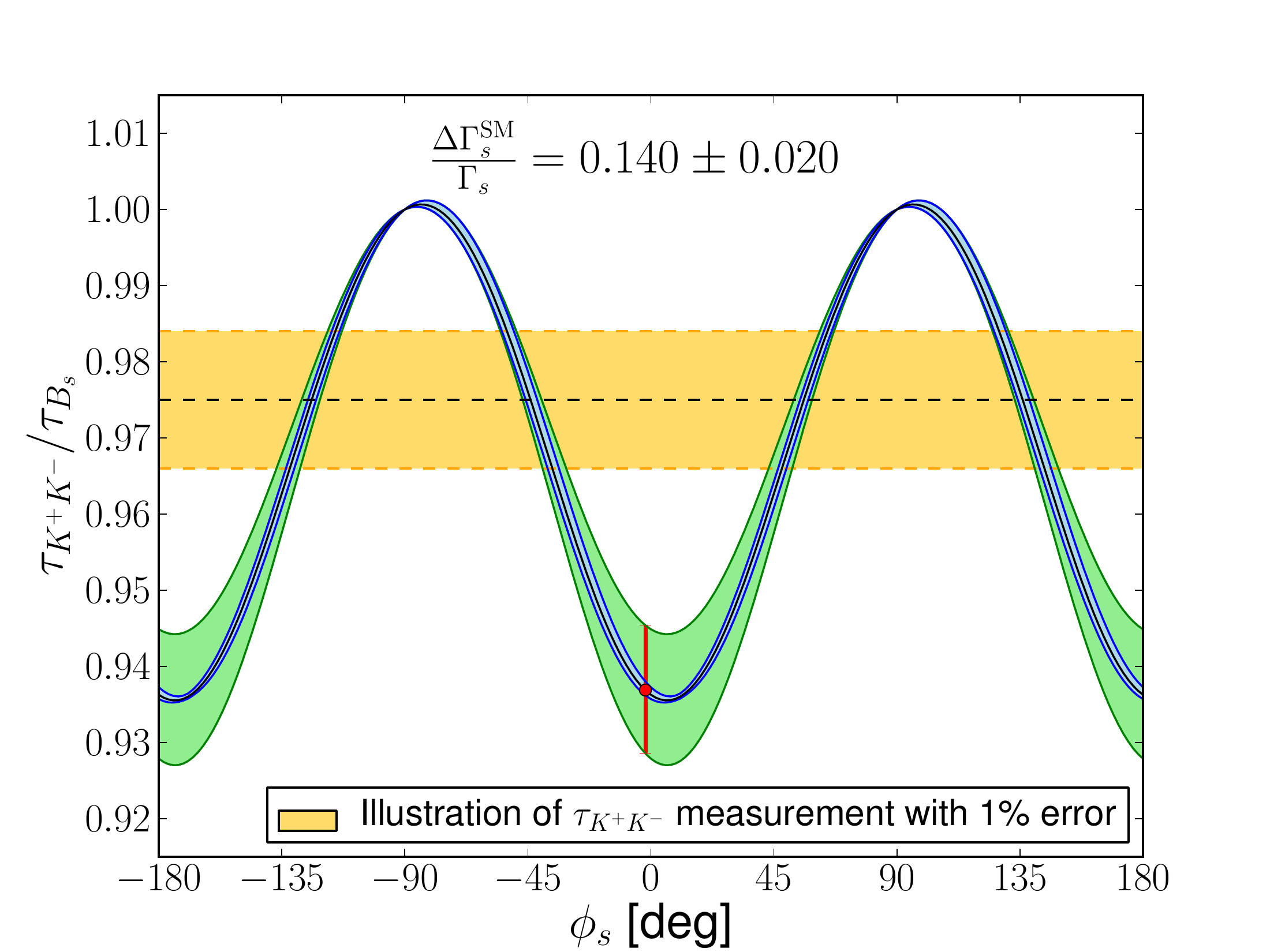}
	   \label{fig:lifetime}
   }
	\subfigure[]{
	\includegraphics[width=7truecm]{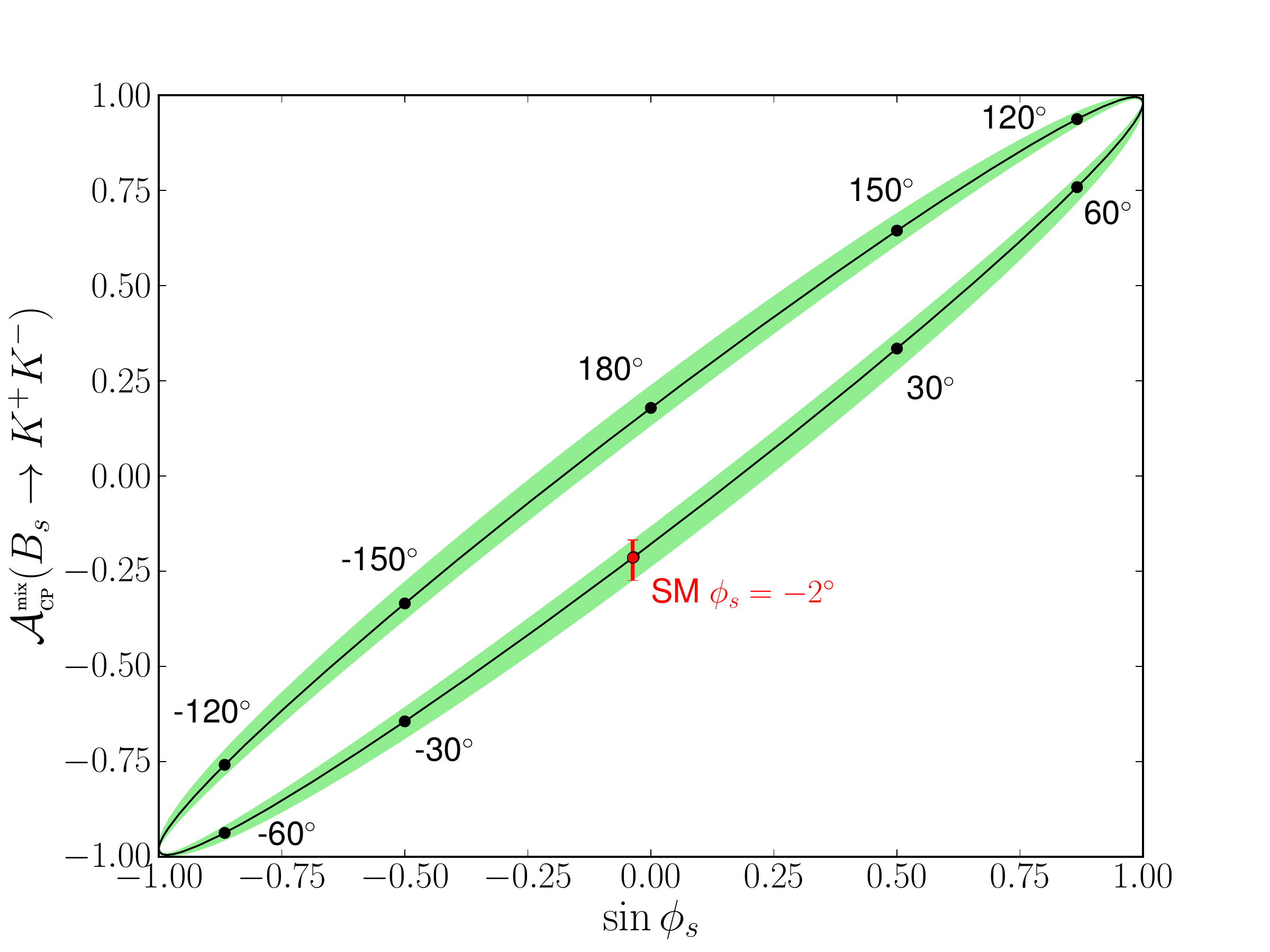} 
	\label{fig:CorrPhiS}
	}
   \caption{Target regions for early LHCb data: {(a)} dependence of 
   $\tau_{K^+K^-}/\tau_{B_s}$ on $\phi_s$, where the narrow band corresponds to the 
   errors of the input quantities and $U$-spin-breaking effects; 
   {(b)} correlation between $\Amix(\BsKK)$ and $\sin\phi_s$, where 
   the error band takes input uncertainties and $U$-spin breaking corrections into account.} 
   \label{fig:target-regions}
\end{figure}

In view of the lack of data on CP violation in $B_s\to K^+K^-$, we use the observable $K$. 
Using then also the CP asymmetries of $B_d\to\pi^+\pi^-$ with $\phi_d$, the $U$-spin
relation (\ref{rel-1}) provides sufficient information to extract $\gamma$ and the hadronic 
parameters. In Fig.~\ref{fig:current}, we illustrate this determination in the $\gamma$--$d$ 
plane by showing the error bands and confidence regions of a $\chi^2$ fit to the current data.
As discussed in Refs.~\cite{RF-BsKK,RF-Bhh} (see also Section~\ref{sec:mixed}), 
the twofold ambiguity can be resolved, thereby leaving us with 
\begin{equation}\label{gam-res}
\gamma = \left(68.3^{+4.8}_{-5.7}|_{\rm input}\mbox{}^{+5.0}_{-3.7}|_\xi\mbox{}^{+0.1}_{-0.2}|_{\Delta\theta}\right)^\circ.
\end{equation}
Here we have also included the impact of possible $U$-spin-breaking corrections, which we
parameterize as $\xi\equiv d'/d=1\pm0.15$ and $\Delta\theta\equiv\theta'-\theta=\pm 20^\circ$.

This result is in excellent agreement with the fits of the UT, giving  
$\gamma=(67.2^{+3.9}_{-3.9})^\circ$ (CKMfitter Collaboration) and $(69.6\pm3.1)^\circ$ 
(UTfit  Collaboration). Consequently, large CP-violating New-Physics (NP) effects at 
the amplitude level are already excluded by the current data. 
However, $B_s\to K^+K^-$ offers sensitive probes for CP-violating NP
contributions to $B^0_s$--$\bar B^0_s$ mixing through its effective lifetime 
\begin{equation}
 \tau_{K^+ K^-} \equiv \frac{\int^\infty_0 t\ \left[\Gamma(B^0_s(t)\to K^+K^-)
+\Gamma(\bar B^0_s(t)\to K^+K^-)\right] dt}
  {\int^\infty_0 \left[\Gamma(B^0_s(t)\to K^+K^-)
+\Gamma(\bar B^0_s(t)\to K^+K^-)\right] dt}
\end{equation}
and the mixing-induced CP asymmetry 
${\cal A}_{\rm CP}^{\rm mix}(B_s\to K^+K^-)$ \cite{FK}. These observables are also 
particularly interesting for improved measurements at the Tevatron and  the early data 
taking at LHCb. Assuming the SM expressions in (\ref{ampl}), we obtain 
the target regions shown in Fig.~\ref{fig:target-regions} as functions of the 
$B^0_s$--$\bar B^0_s$ mixing phase $\phi_s$. 

Once the CP-violating observables of $B^0_s\to K^+K^-$ have been measured, 
an optimized determination of $\gamma$ is possible through the corresponding contour
in the $\gamma$--$d'$ plane \cite{RF-BsKK}. In particular, $K$ is then no longer needed, 
which removes our dependence on the theoretically determined parameter $|{\cal C'}/{\cal C}|$.
In Fig.~\ref{fig:optimal}, we give the projected region for the contour coming from the 
CP asymmetries of $B_s\to K^+ K^-$ that is compatible with the analysis shown in Fig.~\ref{fig:current}.

\begin{figure}[t]
	\subfigure[]{
		\includegraphics[width=7truecm]{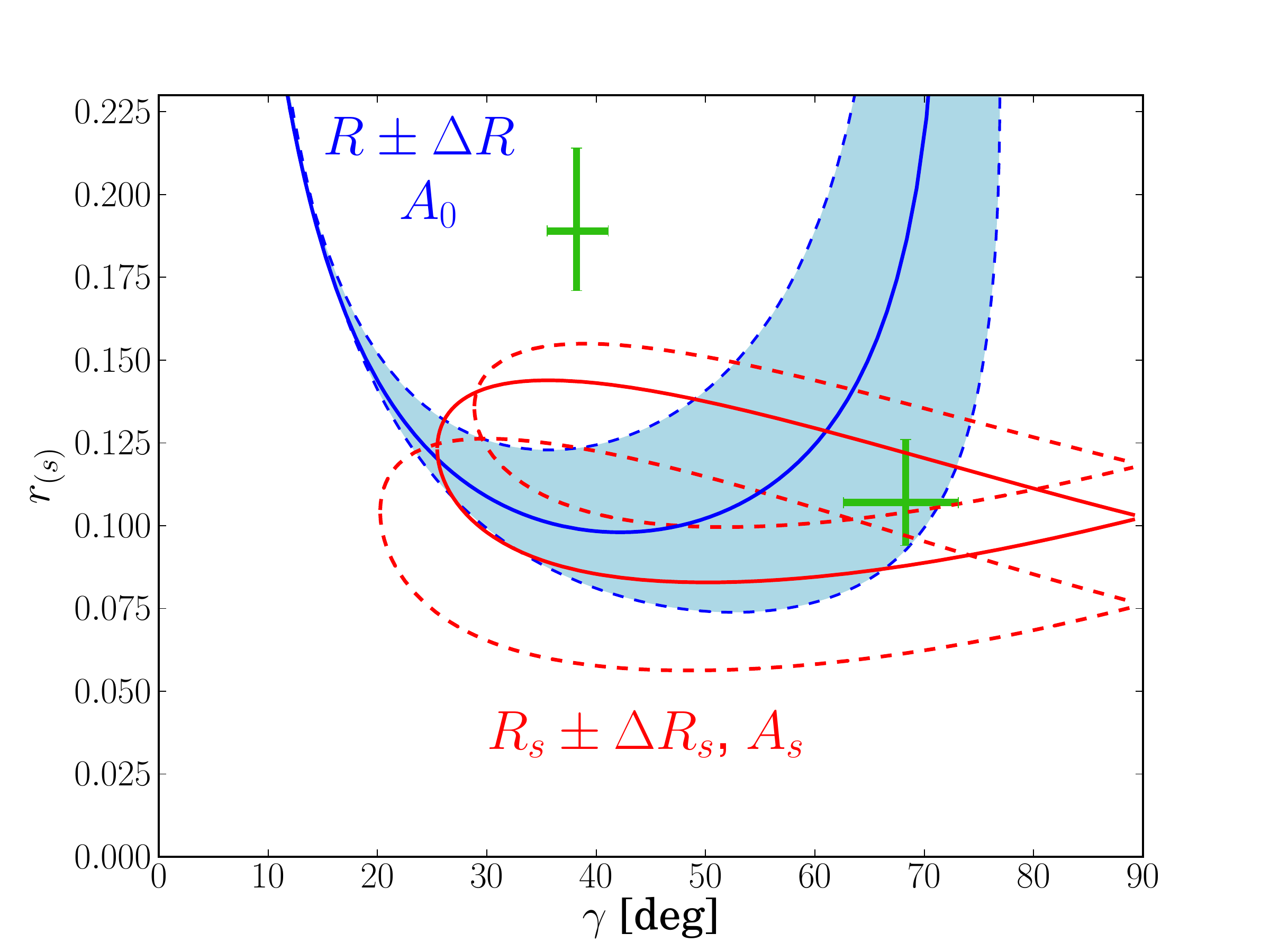}
	}
	\subfigure[]{
		\includegraphics[width=7truecm]{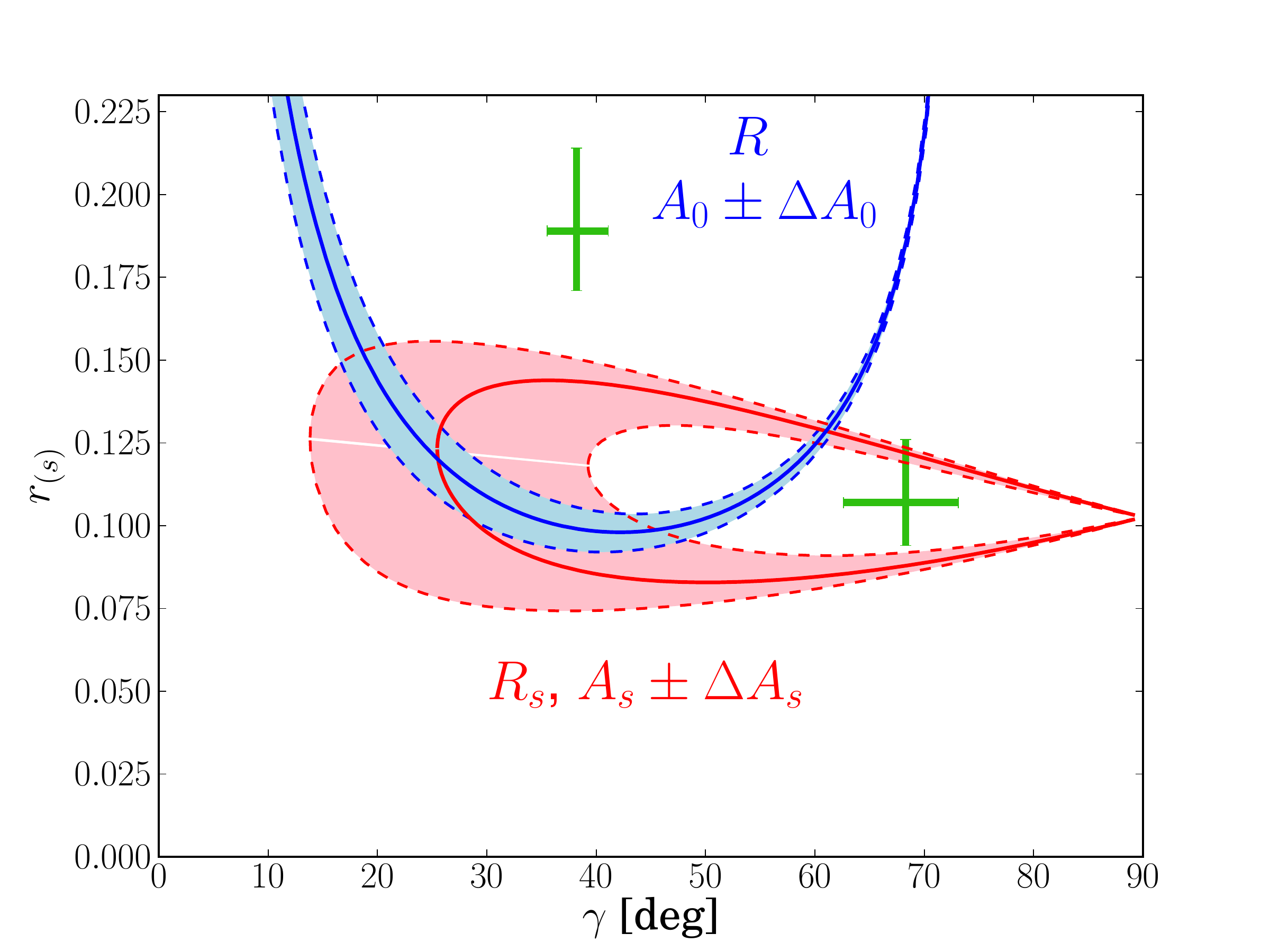} 
	}
    \caption{The current situation of the contours in the $\gamma-r_{(s)}$ plane resulting from the 
   $B_d\to \pi^{\mp}K^{\pm}$, $B_s\to \pi^{\pm} K^{\mp}$ strategy: (a) $1\,\sigma$ error bands
   for $R_{(s)}$; (b) $1\,\sigma$ error bands for $A_{0(s)}$. The two error 
   crosses correspond to the twofold solution in Fig.~\ref{fig:current}.}\label{fig:PiK}
\end{figure}

\section{\boldmath The $B_d\to\pi^\mp K^\pm$, $B_s\to \pi^\pm K^\mp$ System\unboldmath}
\label{sec:mixed}
An alternative for determining $\gamma$ is given by the $U$-spin-related decays 
$B_d\to \pi^{\mp}K^{\pm}$ and $B_s\to \pi^{\pm} K^{\mp}$, which have amplitudes of
the following structure \cite{GR}:
\begin{equation}
A(B_d^0\to\pi^-K^+)=-P \left[1-r\,e^{i\delta}e^{i\gamma}\right], 
A(B_s^0\to \pi^+ K^-)=P_s\sqrt{\epsilon} \left[1+ r_s e^{i\delta_s}e^{i\gamma}/\epsilon\right],
\end{equation}
where $P_{(s)}$ and $r_{(s)}e^{i\delta_{(s)}}$ are CP-conserving hadronic parameters that describe penguin amplitudes and the ratio of trees to penguins, respectively. An unbroken 
$U$-spin symmetry would imply $r_s=r$, $\delta_s=\delta$ and $|P_s/P|=1$. For the extraction
of $\gamma$, the overall normalization $P$ has to be fixed, which can be done through
$B^+\to\pi^+K^0$, neglecting colour-suppressed electroweak penguins and a doubly
Cabibbo-suppressed correction, which is found from $B^+\to K^+ \bar K^0$ data to play a 
minor role \cite{RF-Bhh}. 

It is useful to introduce
\begin{equation}
R\sim\frac{\tau_{B^+}}{\tau_{B_d}}
\left[\frac{\mbox{BR}(B_d\to\pi^\mp K^\pm)}{\mbox{BR}(B^\pm\to\pi^\pm K)}
\right]=1-2r\cos\delta\cos\gamma+r^2\stackrel{\rm exp}{=}0.902\pm0.049.
\end{equation}
This ratio of branching ratios can be converted into the following bound  \cite{FM}:
\begin{equation}
\sin^2\gamma \leq R \, \Rightarrow \, 
\gamma\leq \left(71.8^{+5.4}_{-4.3}\right)^\circ,
\end{equation}
which is nicely consistent with (\ref{gam-res}). Further information on $\gamma$ can be
obtained from 
$A_0\equiv{\cal A}_{\rm CP}^{\rm dir}(B_d\to\pi^\mp K^\pm)R=2r\sin\delta\sin\gamma$.
If we replace
$B_d\to\pi^\mp K^\pm$ through $B_s\to\pi^\pm K^\mp$, we can define -- in analogy to
$R$ and $A_0$ -- observables $R_s=0.250^{+0.065}_{-0.088}$ 
\cite{CDF-BspiK}, which depends -- in contrast to 
$R$ -- on a form-factor ratio (taken from Ref.~\cite{DuMe}), and $A_s$. 
The resulting situation in 
the $\gamma$--$r_{(s)}$ plane shown in Fig.~\ref{fig:PiK} is very similar to that in 
Ref.~\cite{RF-Bhh}, where a much more detailed discussion can be found. 
Because of the $\mbox{sgn}(\cos\delta_s)=\mbox{sgn}(\cos\delta)=1$ 
constraint, only the lower branches of the $\gamma$--$r_s$ contours are 
effective, so that we obtain $24^\circ \leq \gamma \leq 71^\circ$ 
with $0.07 \leq r_{(s)} \leq 0.13$. Consequently, the situation is not as fortunate as in 
the case of $B_d\to\pi^+\pi^-$, $B_s\to K^+K^-$.

Using the dictionary $r e^{i\delta}=e^{i(\pi-\theta)}\epsilon /d$, we obtain the two error 
crosses in Fig.~\ref{fig:PiK}, corresponding to the two solutions in Fig.~\ref{fig:current}.
We can clearly see that the solution around $\gamma\sim 38^\circ$ is excluded. We can
also perform a variety of internal consistency checks of $U$-spin-breaking corrections,
which do not indicate any significant non-factorizable effects within the current
errors \cite{RF-Bhh}.

We look forward to improved measurements by CDF at the Tevatron and the first LHCb data
on the decays considered above!


\end{document}